\documentclass[11pt]{scrartcl}
\usepackage{latexsym,amssymb,bbm,newlfont}
\usepackage{natbib}
\def\aycite{\citet}
\def\ynpcite{\citeyear}
\def\ycite{\citeyearpar}
\def\npcite{\citealt}

\def\be{\begin{equation}}
\def\ee{\end{equation}}

\def\ket#1{ | #1 \rangle }

\def\op#1{ \langle #1 \rangle }
\def\braket#1#2{ \langle #1 | #2 \rangle }
\def\braopket#1#2#3{ \langle #1 | #2 | #3 \rangle }

\def\Bbraket#1#2{ \Big\langle #1 \Big| #2 \Big\rangle }

\def\Hil{\cal H}    
\def\setB{\mathbbm B}
\def\setC{\mathbbm C}
\def\setE{\mathbbm E}
\def\setF{\mathbbm F}
\def\setP{\mathbbm P}
\def\setR{\mathbbm R}
\def\setS{\mathbb S}
\def\setZ{\mathbbm Z}

\begin{document}
\sloppy

\thispagestyle{empty}

\vspace*{-10mm}
\noindent{\emph{Penultimate version, finally to appear in \emph{Studies in History and Philosophy of Modern Physics}}}

\vspace*{20mm}

\noindent{\huge\textbf{\textsf{Berry phase and quantum structure}}}

\vspace*{10mm}

\hspace*{5mm}\parbox{110mm}{
{\large \textbf{\textsf{Holger Lyre}}}\\
\small
Philosophy Department\\
University of Magdeburg\\
Email: lyre@ovgu.de

\vspace*{10mm}

August 2014
}

\vspace*{11mm}

\noindent\textbf{\textsf{Abstract.}}
The paper aims to spell out the relevance of the Berry phase in view of the question
what the minimal mathematical structure is that accounts for all observable quantum phenomena.
The question is both of conceptual and of ontological interest.
While common wisdom tells us that the quantum structure is represented
by the structure of the projective Hilbert space,
the appropriate structure rich enough to account for the Berry phase
is the U(1) bundle over that projective space.
The Berry phase is ultimately rooted in the curvature of this quantum bundle,
it cannot be traced back to the Hamiltonian dynamics alone.
This motivates the ontological claim in the final part of the paper that,
if one strives for a realistic understanding of quantum theory including the Berry phase,
one should adopt a form of ontic structural realism.

\vspace{14mm}


\section{Introduction}

What is the structure of quantum theory? More precisely: what is the minimal structure
that is necessary and sufficient to account for all quantum phenomena?
Conventional wisdom tells us that it is the structure of a non-commutative algebra
of observables represented by self-adjoint linear operators on a Hilbert space.
Since, moreover, the representative of a state is given by a ray in Hilbert space,
the state space even reduces to the projective Hilbert space.
In this paper I argue that this answer is insufficient in so far as
it misses an important element of structure known as the \emph{Berry phase}.
The Berry phase has considerable observable impact within a wide range of quantum phenomena.
To understand its true nature one must scrutinize more rigorously
the state space structure of quantum theory.
It turns out that the appropriate structure to cover the Berry phase is
a U(1) fiber bundle over the projective Hilbert space.

Albeit of central importance in quantum physics, philosophers of physics have largely neglected
the Berry phase or, more generally, geometric and topological phases as a special topic.
One notable exception is \aycite{batterman2003}.%
\footnote{Another exception is \aycite{brown1999}, see also \aycite{sjoeqvist+etal97}.}
But while Batterman thinks that the general lesson of this topic is rather epistemological,
the focus of the present paper is on ontology. I will argue that,
since the Berry phase is ultimately rooted in the state space structure of quantum theory itself,
it is precisely the above-mentioned bundle structure that must be considered as real.
This provides a direct link to structural realism as a doctrine that best captures
the ontological commitments of the physics of Berry's geometric phase.

Basically, the paper consists of two chief claims:
The first is that in order to account for \emph{all} observable phenomena in connection
with a given quantum system the projective Hilbert space structure is insufficient.
It is the U(1) fiber bundle over the projective Hilbert space that is sufficient,
I call it the \emph{quantum bundle}.
The second claim stresses the ontological commitment of the first claim,
namely that the quantum bundle should be considered a real physical structure of nature.
The main reason for this is that, as our analysis aims to show, the Berry phase
cannot be traced back to the underlying Hamiltonian dynamics alone.
It can neither be traced back to the existence of forces,
nor to particular causal mechanisms, nor to the geometry or topology of spacetime.
As a genuine quantum holonomy, the Berry phase is ultimately rooted
in the quantum bundle structure itself. It is an outspring of the curvature of the
quantum bundle, which leads to the ontological conclusion that Berry's phase
supports ontic structural realism about the quantum bundle structure.

The paper is organized as follows:
Section 2 starts with a portrayal of Berry's original derivation of the geometric phase.
It was a major discovery, but it still masks the deeper nature of the phenomenon
which only comes to the fore if the underlying fiber bundle structure is analyzed,
as done in section 3.
Berry's phase turns out as a holonomy of the quantum bundle.
This will further be discussed by distinguishing various types of holonomies in section 4
with the result that Berry's phase is the sole instance of a geometric quantum phase.
Section 5 introduces to what I dub ``Batterman's challenge'':
the quest for an explanation of the ubiquity of holonomy phenomena in physics
and their overall geometric understanding.
\emph{Pace} Batterman I will argue that a mere instrumental understanding
of the physics of such phenomena is dissatisfying and that we should strive
for an ontological picture. I conclude that, in the absence of a sufficient
dynamical-cum-spacetime picture, Berry's geometric quantum phase gives us reason
to consider the quantum bundle structure as an ontologically robust feature of the world.
I end in section 6 with concluding remarks about a structural realist perspective
regarding this structure.


\section{Berry's geometric quantum phase}
\label{berry}

Consider a quantum system that undergoes a cyclic evolution
such that final and initial states are identical.
In general, the system will acquire a `memory' of its evolution in terms of the
so-called \emph{geometric phase} of the wave-function.
This leads to an observable phase shift.
The idea was anticipated by Pancharatnam \ycite{pancharatnam56},
but was fully noted by Michael Berry only \citep{berry84}.
It is therefore widely known as the \emph{Berry phase}
(informal introductions to the geometric phase
can be found in \npcite{berry90b} and \npcite{anandan92}).

For a derivation of the geometric phase along the lines of Berry \ycite{berry84}%
\footnote{For textbook treatments see \cite{boehm_etal2003} and \cite{chruscinski+jamiolkowski2004}.}
consider a system with a Hamiltonian $\hat{H_R}$ that depends on a set of parameters $\vec{R}=(R_1, R_2, R_3 ...)$.
Assume that $\psi_{n,R}(t)$ is a basis of eigenstates of $\hat{H_R}$
with non-degenerate eigenvalues $E_n$ obeying $\hat{H_R} \ket{\psi_{n,R}} = E_n \ket{\psi_{n,R}}$.
We are interested in an adiabatic evolution along a circuit $C$ with period $T$
such that $\vec{R}(0) = \vec{R}(T)$ holds. If the system is initially in state $\psi_{n,R}(0)$,
the adiabadicity requirement guarantees that the system will not jump into
another eigenstate but that later states $e^{i \phi_n(t)} \psi_{n,R}(t)$
will differ by a phase $\phi$ only.
Berry discovered that $\phi$ will generally consist of two parts $\phi_n(t)=\delta_n(t)+\gamma_n(t)$.
This can be seen by plugging $\tilde\psi_{n,R}(t)= e^{i \phi_n(t)} \psi_{n,R}(t)$
into the time-dependent Schr\"{o}dinger equation $i\hbar \frac{d}{dt}\tilde\psi(t)= \hat{H_R} \tilde\psi(t)$
and projecting onto $\tilde\psi_{n,R}(t)$:
\be
\frac{d}{dt} \, \phi_n(t) =
i \Bbraket{ \psi_{n,\vec{R}} }{ \vec{\nabla}_{\!\!\vec{R}} \ \psi_{n,\vec{R}} } \ \frac{d}{dt} \, \vec{R}(t)
- \frac{1}{\hbar} E_n .
\ee
Integration over $t$ yields the total phase consisting of
the integral $\delta_n(t) = - \frac{1}{\hbar} \int E_n(t) dt$,
called the dynamical phase representing the time evolution,
and the ``geometrical'' phase $\gamma_n(t)$,
which after a closed evolution yields
\be
\label{berryphase}
\gamma_n(C) = i \oint_C \braket{ \phi_{n,\vec{R}} }{ \vec{\nabla}_{\!\!\vec{R}} \ \phi_{n,\vec{R}} } \ d\vec{R}.
\ee

The derivation of Berry's geometrical phase within a pure quantum framework
already indicates that we are dealing with a genuine quantum effect.
Moreover, as Berry already noted, $\gamma_n$, is non-integrable,
it cannot be written as a function of $\vec{R}$, but seems to depend on the particular path
in $\vec{R}$-space -- which is, curiously, an abstract mathematical parameter space
(we will return to this point later).

Geometric quantum phases play a role in physics in a startling variety of phenomena
ranging from optical physics, interference effects, crystal physics,
nuclear magnetic resonance and others (\aycite{shapere+wilczek89} give
a comprehensive overview by providing a collection of relevant papers).
The general setting of the observation of such phase is by interference
of a cycled system with another uncycled system.
A standard example is for instance a spin-$\frac{1}{2}$-particle in an adiabatically
rotating magnetic field such that the spin aligns with the field.
After a loop in spinor space the particle wave-function has acquired a complex phase.
Another well-known example is Pancharatnam's already mentioned discovery of the phase shift
of polarized light that is taken through a cycle of polarization states \citep{pancharatnam56}.
It is, in retrospect, astonishing that such a broad class of phenomena
was discovered as late as in the 1980's only, more than half a century
since quantum theory had been fully developed.


\section{The fiber bundle structure of the quantum phase}
\label{deeper}

Soon after Berry's seminal work the geometric quantum phase has been recognized
in its full generality.
It is, as \cite{simon83} has first pointed out (cf. also \npcite{page87}),
appropriately captured in the framework of fiber bundles and may,
as Aharonov and Anandan \ycite{aharonov+anandan87} were able to show,
be generalized to non-adiabatic cases as well.
From these two insights it turns out that the crucial structure from which the Berry phase
originates is given by a $U(1)$-fiber bundle over the projective Hilbert space $\Hil P$.
Call this the \emph{quantum bundle}:
\be
\label{corestruc}
U(1) \ \to \ \Hil \ \to \ {\Hil} P \qquad \mbox{or} \qquad \setP({\Hil} P, U(1) ) .
\ee
The quantum bundle defines a projection $\pi\! : \ \Hil \to {\Hil} P$ from the total bundle space,
the Hilbert space $\Hil$, to the base space $\Hil P$.

In the following subsection, a few remarks about projective spaces shall first be made.
Next, the derivation of the Berry phase from the quantum bundle structure will be shown.
It turns out that the Berry phase is in fact a holonomy that does not depend on the
trajectory in parameter space but in the projective Hilbert space.
This observation has crucial repercussions for our line of argument.

\subsection{Projective spaces}

In the case of $\setR^3$ the corresponding projective space $\setR P^2$
is the set of all lines in $\setR^3$ passing through the origin.
Note that every projection line meets a sphere centered in the origin exactly twice at antipodal points.
Hence, $\setR P^2$ is the set of points on the unit sphere $\setS^2$ modulo an identification of antipodal points.
Alternatively, it is the set of equivalence classes of points $(x, y, z) \in \setR^3 \backslash \{(0, 0, 0)\}$,
where $(x, y, z)=(\lambda x, \lambda y, \lambda z)$ for a non-zero $\lambda \in \setR$.
It follows that $\setR P^2$ is diffeomorphic to $\setS^2 / \setZ_2$,
where $\setZ_2$ is the cyclic group of order 2. More generally we get
\be
\setR P^n \simeq \setS^n / \setZ_2 .
\ee

The connection between projective spaces and quantum theory rests on the long known fact
that the representative of a quantum state is not a vector but a ray in Hilbert space,
i.e.\ a one-dimensional subspace or equivalence class $\psi \sim \lambda \psi$
with $\psi \in \Hil$ and $\lambda \in \setC$.
The principal argument for this is that the transformation $\psi \to \lambda \psi$ leaves
the expectation values $\op{\hat O} = \frac{\braopket{\psi}{\hat O}{\psi}}{\braket{\psi}{\psi}}$
of all observables $\hat O$ invariant.
Hence, the projective Hilbert space $\Hil P$ as the set of rays is considered to be
the state space of quantum theory. But this also leaves us with a puzzle.
For even if we restrict ourselves to normalized state vectors by the requirement $\braket{\psi}{\psi}=1$,
as we standardly do, two states that differ by a phase $\psi'=e^{i \varphi} \psi$ are equivalent.
The phase difference drops off and phases seem to play no role in quantum theory any longer
-- at least if we follow the above line of argument.
But not only does this render the original theory puzzlingly redundant,
it also leaves no room to account for the Berry phase.
The answer to this puzzle is that neither $\Hil$ nor $\Hil P$ gives us the necessary
and sufficient structure to account for \emph{all} observable phenomena.
We must rather consider the bundle structure (\ref{corestruc}), a structure that
sits, so to speak, right between $\Hil$ and $\Hil P$ (in the sense that,
considered as sets, we have $\Hil P \subset \setP({\Hil} P, U(1) ) \subset \Hil$).
It is precisely this structure that accounts for the Berry phase.

\subsection{The quantum bundle $\setP({\Hil} P, U(1) )$, its connection and holonomy}

Consider a finite ${(n+1)}$-dimensional quantum system with a Hilbert space ${\Hil} = \setC^{n+1}$.
Standard normalization of the states defines the sphere $\setS^{2n+1}$.
The corresponding projective space is $\setC P^{n} \simeq \setS^{2n+1}/\setS^1$.
The quantum theory of an $(n+1)$-dimensional system is then represented by the bundle
\be
U(1) \ \to \ \setS^{2n+1} \ \to \ \setC P^n \simeq \setS^{2n+1}/\setS^1.
\ee
The generalization of this result leads to the bundle structure (\ref{corestruc})
already introduced in the beginning of this section.

It will be sufficient to sketch the derivation here,
the reader is referred to \aycite{boehm_etal91} and \ycite{boehm_etal2003} for details.
A bundle connection serves to define the notion of parallel transport in a bundle
and thus serves to identify neighboring fibers.
Given the tangent space $T \setE$ of a bundle $\setE$ the connection is likewise a rule to define
the horizontal subspace in the decomposition $T_u\setE = V_u\setE + H_u \setE$ at any point $u \in \setE$.
Consider now tangent vectors $\dot\varphi \in T \setE$ to the curve $\varphi \in \setE$ which can be decomposed
via the scalar product into
\be
\label{varphi_curve}
\ket{\dot\varphi} = \ket{v_\varphi} + \ket{h_\varphi} = \braket{\varphi}{\dot\varphi} \ \ket{\varphi} + \ket{h_\varphi} .
\ee
Note that two Hilbert space vectors that differ by a phase belong to the same fiber and point into
the vertical fiber direction, hence the horizontal component satisfies $\braket{\varphi}{h_\varphi}=0$.
For the curve $\varphi$ to be a horizontal lift, the vertical part of (\ref{varphi_curve}) must vanish
\be
\label{hl_eq}
\braket{\varphi}{\dot\varphi} = 0 .
\ee
A bundle holonomy is given by the horizontal lift of a closed curve $\ket{\phi}$ in base space
with $\ket{\phi (0)}=\ket{\phi (T)}$ and $\ket{\varphi}=e^{ig(t)}\ket{\phi}$.
Inserting this into (\ref{hl_eq}) and integrating over $t$ yields
\be
\gamma = g(T)-g(0)=i \oint_0^T \braket{\phi}{\dot\phi} \ dt .
\ee

We have, in full generality and without any recourse to an approximation (adiabatic or otherwise),
re-discovered the quantum phase as the holonomy of the connection of the quantum bundle (\ref{corestruc}).
This connection, known as the \emph{Berry connection}, is simply the scalar product in Hilbert space.

In general, the phase depends of course on the choice of the connection.
It is therefore important to note that \aycite{boehm_etal92} were able to show
that the Berry connection is unique if we require invariance under the group of unitary transformations.
Because of its direct connection to the conservation of probability,
the unitary group can indeed be considered the canonical transformation group in quantum physics.
The splitting between the dynamical and the geometrical phase is in this sense non-arbitrary,
$\delta$ and $\gamma$ have independent physical meaning.

Important for the purposes of this paper is now the observation that unlike
the derivation in section \ref{berry} the derivation of the generalized Berry-Aharonov-Anandan phase
does not depend on the structure of the parameter space
but on the projective Hilbert space.
As Aharonov and Anandan (\ynpcite{aharonov+anandan87}, p.~1594) put it:
\emph{``we regard $\gamma$ as a geometric phase associated
with a closed curve in the projective Hilbert space and not
the parameter space, even in the special case considered by Berry.''}
Moreover, $\gamma$ is reparametrization invariant and therefore
independent of the rate or speed at which the evolution has occurred.
In fact, an infinite number of Hamiltonians will generate motions in $\Hil$
which project onto the same curve in the projective Hilbert space
and which therefore reproduce the same quantum phase.

In section \ref{holonomies}, various types of holonomies will be distinguished and
the particular type of the Berry phase as a geometric quantum phase will be explored.
First, however, a short remark regarding the logico-algebraic structure of quantum theory:
The discovery of the Berry phase and its deeper geometric nature has, it seems, repercussions
for quantum theoretical programs that focus on the projective Hilbert space structure $\Hil P$ alone.
This is seldom explicitly stated, but as we have seen, the necessary and sufficient structure
to account for the Berry phase isn't $\Hil P$ but rather the $U(1)$ bundle over $\Hil P$.
The structure of $\Hil P$ is most prominently studied in quantum logic,
where one combines projective geometry with lattice theory (more precisely, the set of closed subspaces
of the projective Hilbert space determines a special case of a non-Boolean orthomodular lattice).
But the foregoing analysis sheds doubts on the prospects of quantum logic (or other programs
that focus on the projective structure only) to be sufficient to account for the the Berry phase and,
hence, for \emph{all} quantum phenomena. The point is that the Berry phase is not represented
by a self-adjoint operator, but as a holonomy of $\setP({\Hil} P, U(1) )$.
It is the quantum bundle structure (\ref{corestruc}) that must be considered
the appropriate quantum structure.

On the other hand, to say that the bundle structure (\ref{corestruc}) is fundamental
in quantum theory seems to create a tension in view of the standard operations in $\Hil$
such as the superposition of vectors and the action of operators, since they are defined on $\Hil$
as a vector space (an algebraic rather than a geometric structure).
The tension, however, vanishes by the natural requirement that all such operations
preserve the projection $\pi\! : \ \Hil \to {\Hil} P$.
The transition probability for two states (rays), for instance, is defined by the `angle'
in the complex projective space that is fiber preserving (cf.\ \npcite{boya+sudarshan89}).


\section{The ubiquity of holonomies}
\label{holonomies}

How does Berry's holonomy relate to other holonomies in physics?
The ubiquity of physical holonomy phenomena calls for careful conceptual distinctions.
Geometrical and topological effects can already be found in classical physics.
We may distinguish between geometric/topological holonomies on the one hand and
classical/quantum holonomies on the other as well as the Aharonov-Bohm holonomy as a special case.

\subsection{Geometric versus topological holonomies}

A familiar example of a classical holonomy is the parallel transport of a vector over a sphere
along a loop. Consider the case where a vector is tangent to the sphere at the North pole.
The vector may now be parallel transported along a great circle down to the equator,
then further transported along the equator (say, a quarter of the whole equator)
and taken back to the North pole along another great circle.
Compared to its initial state the vector will now point into a different direction:
it will have acquired a holonomy of $\pi/2$.

Now consider the case of a M\"{o}bius strip.
Take again a vector, but this time perpendicular to the surface.
Transport it along the M\"{o}bius strip and after one circuit its direction
will be reversed (and it will return to its initial position after another circuit).
We have again produced a holonomy.
But the two cases are different. In the case of the sphere the particular route is important
and the holonomy will in general vary by a variation of the loop (suppose we had
transported the vector along half of the equator: we would have produced a holonomy of $\pi$).
In the case of the M\"{o}bius strip only multiples of a full circuit count and will solely
produce two discrete holonomy outcomes: reversed (-1) or non-reversed (+1).
Indeed, the holonomy in the case of the sphere or, more generally, in the case
of curved spaces can be used to measure the curvature of the underlying space.
The parallel transport is non-integrable, i.e. it depends on the particular
path and does not depend on initial and final points only.%
\footnote{Strictly speaking this must be called an anholonomy, since it relates
to the distinction between integrable and non-integrable, anholonomic constraints
in classical mechanics, but it is common language use in differential geometry
to call it a holonomy as well.}
The holonomy in the case of curved spaces can be traced back to the
curvature properties of the underlying space. We may say that
\emph{it detects the geometric properties of that space}
and can properly be called a \textit{geometric holonomy}.
By way of contrast, the holonomy of the M\"{o}bius strip detects a topological feature
of the underlying space and is therefore a \textit{topological holonomy}.

The difference can very well be expressed in the language of fiber bundles.
Our first scenario defines a tangent bundle over the sphere, while the M\"{o}bius strip
represents a line bundle over the circle. Both bundles are non-trivial.
In the general case of a tangent bundle on a curved space the bundle could either be
trivial or non-trivial depending on the topological nature of the base space.
Generally, a fiber bundle with total space $\setE$, base space $\setB$ and fiber $\setF$
is trivial if it allows for a global section
(think of the global structure as a cross product $\setE = \setB \times \setF$).
Obviously, the bundle is non-trivial if the base space has a non-trivial topology
as in the cases of sphere and circle.
More crucial for our purposes is the question whether the bundle has
a vanishing or non-vanishing curvature tensor. In the former case,
the connection, the derivative of which gives the curvature, is flat,
while in the latter case the bundle has a non-flat connection.
The tangent bundle over the sphere is a bundle with a non-flat connection.
The M\"{o}bius strip, however, has a flat connection.
Its holonomies are due to the non-trivial topology of the base space.

\subsection{Classical versus quantum geometric holonomies}

After highlighting the distinction between geometric and topological holonomies,
we should consider the distinction between classical and quantum cases.
The cases of parallel vector transport on the sphere or through a M\"{o}bius strip
are classical cases. Another classical example is the following:
classical oscillators that undergo, under special circumstances, a cyclic evolution
exhibit a holonomy that is known as the Hannay angle \citep{hannay85}.
This has often be pointed out as a classical analogue of the geometric quantum phase.
A special and probably best known application is Foucault's pendulum.
Here, the change of direction of the plane of swing can be described as a case of parallel transport
by modeling the precession by the projection of the angular velocity of the Earth onto its normal direction.
Foucault's pendulum is therefore a geometric holonomy in the same sense as the
parallel transport of a vector on a sphere. By way of contrast, Berry's phase is a quantum phase
in the sense that it is picked up by a wave-function. So it restricts to phases of quantum systems.
At the same time, and as already pointed out in section \ref{deeper},
it depends on the cyclic evolution in the projective Hilbert space.
It is in this sense a geometric quantum phase or, as it were, a ``projective-geometric'' quantum phase
because of the fact that infinitely many Hamiltonians will generate motions in $\Hil$
projecting onto the same curve in $\Hil P$ and producing the same phase.

\subsection{The Aharonov-Bohm effect}

The Aharonov-Bohm effect \citep{aharonov+bohm59} deserves special attention
since it is often considered to be an instance of the Berry phase.
The scenario is very well known:
a split electron beam passes around a solenoid in which a magnetic field is confined.
The region outside the solenoid is field-free, but nevertheless a shift in the interference pattern
on a screen behind the solenoid can be observed upon alteration of the magnetic field.
The phase shift can be calculated from the loop integral over the potential,
which---due to Stokes' theorem---relates to the magnetic flux
$$\Delta \chi
= q \oint_{\cal C} \vec A \ d \vec r
= q \int_{\cal S} \vec B \ d \vec s
= q \ \Phi_{mag}.$$
The AB effect has triggered a host of theoretical considerations and experimental work
(a comprehensive overview can be found in \npcite{peshkin+tonomura89}).

What is the nature of the AB effect? A first observation is that the AB effect
is not of genuine quantum nature, since there exist classical AB scenarios.
As a simple classical example consider the geometry of a cone with flat curvature (i.e.\
a vanishing gravitational field) everywhere except at the apex (which may be smoothed).
Parallel transport on a loop enclosing the apex leads to a holonomy.
The example provides a proper AB setting within classical (field) physics.
The possibility of AB scenarios in both the classical and the quantum regime suggests
that there is another structural feature to which the effect can be traced back.
Common to the magnetic and electric AB effects in QM and gravitational AB cases
is the fact that both underlying interactions, electromagnetism and gravity,
can be described within the framework of gauge theories.
This seems to indicate that the effect is tied to the
underlying gauge theoretic structure (which is closely related to the claim
that the fundamental entities in gauge theories are holonomies;
cf. \npcite{healey2007}, \npcite{lyre2004}).
This also fits to the fact that the AB effect can be generalized
to arbitrary SU(N) gauge groups \citep{wu+yang75}.

Moreover, the AB effect does not depend on the particular shape of the path
if only the region of a non-vanishing gauge field strength is enclosed.
This is the crucial similarity between the original magnetic AB effect
and its classical gravitational analogue -- and it relates to the above mentioned case
of the topological holonomy of the M\"{o}bius strip in the specific sense
that in this case, too, the holonomy does neither depend on the particular path
nor does the path pass a region with non-vanishing curvature.
That is, in both cases the connection is flat. In the AB case, however, it isn't flat throughout.
What is characteristic of the AB scenario is that a region of a non-vanishing gauge field strength
is enclosed (the confined magnetic field or the smoothed apex).
It has, however, nothing to do with a non-trivial topology of the underlying base space.
In both the original AB setting and in the case of the cone
the base space is topologically well behaved and trivial.
In this sense the AB effect is not a topological effect in the rigorous sense
that it relates to the non-trivial topology of the underlying base space.
The upshot is that the AB effect fits neither into the category of
a pure topological nor a pure geometric holonomy, it depends on the existence
of a confined gauge field. It is thus no instance of the Berry phase.


\section{Batterman's challenge and minimal quantum structure}

We have seen from section \ref{berry} that the Berry phase plays an eminent role
throughout quantum physics and for a large spectrum of phenomena.
Albeit of great experimental and theoretical importance, philosophers of physics
have largely overlooked Berry's quantum phase as a special topic.
Given the experimental evidence this is indeed a serious neglect.
One notable exception is Robert Batterman \ycite{batterman2003}.
Batterman's paper draws attention to the ubiquity of holonomies
by highlighting many important physical and mathematical facts.
His discussion runs over many of the issues we already touched upon in the foregoing sections:
ideas of parallel transport, Pancharatnam's polarization phase, Foucault's pendulum,
the AB effect, fiber bundles, and so on.
As Batterman emphasizes, his focus is on the explanatory value of such geometric phenomena
rather than on ontology. He encounters a challenging puzzle:
\emph{``It turns out that you can explain the physical changes that appear
as a result of ... round trip excursions by appeal to certain purely geometrical features
of the abstract space in which the excursion can be parameterized. This is, \emph{prima facie}, odd.
What sort of role can geometrical/topological features of some abstract space
play in \emph{explaining} and providing \emph{understanding} of ``real'' physical phenomena?''}
(Batterman 2003, p.~527)

Indeed, that sounds odd. If real physical phenomena were to depend on the properties of
some abstract, mathematical parameter space, then we seem to be left with an unfortunate alternative:
either we come to the conclusion that our physical explanation of the phenomena is strictly speaking wrong,
since it doesn't provide us with a full-blown explanation drawing on either causal mechanisms,
forces, or, as we will discuss below, spacetime geometry, or we bite the bullet and come to terms with Platonism.
Batterman -- rather implicitly than explicitly -- seems to endorse the first option.
He acknowledges the ubiquity of geometric and topological phenomena in physics,
and argues that the universality of such phenomena calls for unifying and wide-ranging explanations.
The description in terms of holonomies provides us with such a universal explanation,
but, as Battermain claims, without any need to take it metaphysically serious.
After all, in all those types of explanations certain crucial idealizations and approximations are involved.

However, caution is demanded. Is it really true that idealizations alone account for the ubiquity of holonomies?
Idealizations and approximations are Batterman's special topic
(compare \npcite{batterman2002}, \ynpcite{batterman2010}),
and they obviously play an eminent role in many places in physics and the sciences throughout.
But certainly not \emph{all} explanations in terms of holonomies are explanations which invoke
idealizations or approximations. Holonomies are no mysteries.
In many cases we \emph{are} able to tell a causal-mechanistic story (including physical forces),
other cases allow us to tell a story in terms of the underlying geometry of position space.

Consider again the case of parallel vector transport in curved space.
Obviously, what is responsible for the occurrence of holonomies in such a case
is the curvature of the underlying base space. Likewise the case of the M\"{o}bius strip.
What brings about this type of holonomy is the non-trivial topology of the base space.
We have thus two simple cases, cases of classical holonomies,
for which a non-mysterious story can very easily be told.

By way of contrast, consider Foucault's pendulum. Here, the description in terms
of parallel transport and holonomies is really just an idealized way to describe what's going on.
What brings about the precession of the oscillation plane of Foucault's pendulum?
From the perspective of an earth-bound reference frame, it can be seen as an effect of the Coriolis force.
But since this force is a pseudo-force reflecting the rotation of the earth,
it is the rotation of the earth itself that is ultimately responsible for Foucault's holonomy.
This raises the suspicion that in all cases of classical holonomies the effect can either be
traced back to the influence of forces or has to do with the geometrical or topological
properties of position space itself as the underlying base space.
No mysterious abstract spaces or idealized explanations must be invoked.

A rather peculiar case is the special relativistic case of Thomas precession.%
\footnote{I like to thank Harvey Brown and David Malament for independently pointing this out to me.}
Consider a gyroscope travelling along a loop in flat Minkowski spacetime.
After the loop the gyroscope's axis will not align with its original direction, it has acquired a holonomy.
What brings about this effect? No forces are acting on the gyroscope, nor is position space curved.
Standard treatments make reference to the fact that the \emph{space of velocities}
in special relativity is hyperbolic, and so the parallel transport of a vector changes its direction.
Should we say that the gyroscope detects the geometry of velocity space,
which then brings about the precession effect?

One could put it this way, we should however recognize the fact that, unlike classical Euclidean spacetime,
the geometry of Minkowski spacetime mandates that translations and momenta are non-trivially connected,
and only together form the geometrical symmetry group of special relativity.
Thomas precission can algebraically be understood as an outspring of the Lorentz group
(more precisely, the non-associativity of velocities).
So even if the effect cannot be traced back to the curvature of position space,
it still reflects the geometrical symmetry of Minkowski spacetime.
Again: classical holonomies can either be traced back to forces or to the geometrical structure of spacetime,
or so it seems.%
\footnote{I leave this as a hypothesis. Nothing hinges on its truth for the purposes of this paper,
which ultimately focuses on the quantum Berry phase).}
The real threat comes with quantum holonomies.

What brings about the Berry phase?
The effect is neither brought about by forces nor by the geometry of spacetime.
Take the shift of the wave-function of a spin-$\frac{1}{2}$-particle in a rotating magnetic field.
Of course, the rotation of the magnetic field is a force- or interaction-like process.
But it alone cannot account for Berry's phase. Moreover, as we have already seen,
an infinite number of Hamiltonians may produce the same phase.
The Hamiltonian dynamics leading to the cyclic evolution in $\Hil P$
is at best a necessary, but not a sufficient condition for Berry's phase.
What is crucial, however, is the fact that $\Hil P$ has a special geometry.
So what brings about the effect?
On the basis of the bundle-theoretic derivation in section \ref{deeper}
we are now in the position to give a precise answer to this question:
\emph{It is the curvature of the quantum bundle that brings about Berry's phase.}
Any quantum evolution is constrained by the geometry of (\ref{corestruc}),
the $U(1)$ bundle over the projective Hilbert space.
Berry's quantum phase detects, as it were, the quantum bundle curvature.

This is remarkable, since this is the first instance of a physical holonomy
in our discussion so far that draws on a piece of physical structure that exceeds
dynamical-cum-spacetime structure.
It also becomes clear that the structure of the quantum bundle is the minimal structure
that is necessary and sufficient to account for all quantum phenomena.
The structure of the quantum world cannot be accounted for by the structure of $\Hil P$,
but only by the structure of a $U(1)$ bundle over $\Hil P$.

It is this extra piece of quantum structure that provides us with a third option
regarding Batterman's challenge that goes beyond the alternative of either idealizations
due to a missing dynamical-cum-spacetime picture or Platonism.
The third option is that, since the Berry phase can be rediscovered as a holonomy of the Berry connection,
the existence of that phase is directly rooted in the quantum bundle structure.
Batterman's analysis misses this option since he only refers to Berry's original derivation,
which traces the phase back to parameter space (as shown in section \ref{berry}).
If this were the end of the story, this would indeed be ``odd''.
Berry's phase, however, doesn't live in parameter space, but in the quantum bundle.
It detects the curvature of that bundle, any quantum loop evolution is constrained by it.

But why not consider the quantum bundle structure as an idealization already?
The reason is that even an instrumentalist or operationalist about quantum theory must use
this much structure at minimum to account for all observable quantum phenomena.
He cannot just work with the structure of $\Hil P$,
only the $U(1)$ bundle over $\Hil P$ will do.


\section{Quantum structure and structural realism}

The aim of the paper was to spell out the relevance of the Berry phase
in view of the question what the minimal mathematical structure is that accounts
for all observable quantum phenomena. This aim has been reached by identifying
the quantum bundle (\ref{corestruc}) as the minimally relevant structure.
The paper could stop at this point, but in this final section I want to
go one step beyond a merely conceptual analysis of the structure of
quantum theory and draw possible ontological conclusions from it.
Non-realists will surely not follow, but the scientific realist,
who is willing to make an ontological commitment,
cannot overlook the relevance of the quantum bundle as the minimal quantum structure.
It is the purpose of this final section to spell out a moderate and at the same time
straightforward realist commitment in connection with the above findings.
My recommendation is to be a realist about the quantum bundle structure
reflecting the quantum structure of the physical world.

Let me emphasize that it is not the purpose of this paper
to present arguments in favor of structural realism \emph{per se}.%
\footnote{I've done this elsewhere, e.g.\ Lyre \ynpcite{lyre2004}, \ynpcite{lyre2012};
see also \aycite{french2014} for a recent comprehensive anthology on structural realism.}
I rather want to argue that, if one aims for an ontological picture of
quantum physics including the Berry phase, structural realism appears
to be a promising, if not the most promising option.
Philosophers of Batterman's stripe will certainly not be convinced.
They will stick with the view that sometimes abstract mathematical structures
can play an explanatory role in empirical science---no further comments necessary.
However, in the absence of a missing dynamical-cum-spacetime picture,
I find this highly dissatisfying (as elaborated on in the previous section)
and therefore recommend to look for an ontologically robust interpretation.
The opposite side might insist that it is as odd to consider the quantum bundle
structure as real as to consider Berry's original parameter space as real.
But note that there is a vast number of Berry phase scenarios
with a vast number of particular parameter spaces.
It would be inflationary (and crazy) to consider all of them as ``real.''
Fortunately, there is only one quantum bundle to account for
this vast number of observable effects.
Its structure (basically encoded in the bundle connection) is,
as we've seen, the minimal structure necessary and sufficient
to account for all observable quantum phenomena.
It is precisely this minimality feature that makes it a distinguished piece
of structure and thus a candidate structure for the quantum structural realist.
And there is nothing mysterious about the quantum bundle structure
as compared to other candidate structures, mostly groups and algebras,
that have already been identified by SR proponents by the same rationale.

Let me elaborate on this.
Quite generally, the scientific realist who takes quantum theory serious will acknowledge
that the world is of quantum nature -- which is to say that it has a quantum structure.
That structure is naturally encoded into the quantum theoretical state space:
it tells us how physical states are related to each other.
But as we have seen, the quantum physical state space $\Hil P$ not only encodes
the state transitions in terms of its algebraic structure (as quantum logic would have it),
but also constraints the temporal evolution of quantum states in a subtle
geometrical way such that, in the case of loop evolutions, holonomies occur.
It seems therefore natural to advocate a form of realism about the quantum bundle structure
and, hence, to advocate a form of ontic structural realism (OSR).

To motivate our structural move in a more ``relationalist'' fashion
consider the following:
For the Berry phase to become manifest and observable we need to compare two systems.
Perhaps the peculiarity of the Berry phase lies in the fact that it exists
only as a relation between two systems?
A critic might indeed point out that the effect is manifest as a relative phase only
and that there is nothing special about relative phases in quantum theory.
It is true that relative quantum phases are in principe observable.
Take a quantum beam splitter, e.g.\ a double slit, and consider a phase shifter in one of the beams,
then surely the relative phase becomes observable as a shift in the interference fringe.
But note again that this effect is perfectly understandable in terms of the
(more or less) locally acting phase shifter. This, again, is not the case for the Berry phase.
It is true that it becomes observable as a relative phase between two systems only,
an evolved system and a reference system, but the change in the relative phase or,
more generally, the change in the relation between the two systems is not
brought about by any forces or the like. So, how does Nature do it?
What ontological picture accounts for a world that shows effects like this?

At this point a structuralist move suggests itself.
SR assumes (at least in its moderate version) that on the fundamental level
there are otherwise non-individuated objects with relational properties only.
Such relations are globally orchestrated by the (various) structure(s) of the world.
Structural realists should thus be prepared that Nature behaves such that relational
changes might occur without the existence or influence of locally acting forces.
The Berry phase is thus an outspring of the quantum structure of the world.
More precisely: In a quantum world it is to be expected that relational changes in terms of
Berry's geometrical holonomy will occur, since this holonomy is built into the world's quantum structure.
There is no quantum world without Berry phases. The relational changes spelled out by the Berry phase
are not brought about by forces, they are built into Nature's bottom structure.
From an ontological point of view, we may account best for this effect by assuming an ontological
picture that does not construe the world to consist of a mosaic of local entities,
but rather to consist of global structures.

One might compare the quantum bundle structure to the constraint that permutation symmetry
imposes on the allowed physical states in a many particle state space.
Here, too, the permutation symmetry structure brings about real physical effects,
frequently spelled out in connection with Pauli's exclusion principle.
An impressive instantiation of this is, for instance, the electron degeneracy pressure
that leads to the stability of white dwarfs and neutron stars by preventing them
from collapsing under their own weight. OSR proponents will therefore also take
the permutation symmetry structure as a real ingredient of our physical quantum world.

Note that OSR isn't Platonism, in fact, it is quite the opposite of Platonism.
Of course, fiber bundles, Hilbert spaces, and the like are mathematical entities in the first place.
Ontic \emph{in re}-structural realism is, however, the view
that the basic structures that occur in the mathematical representation of our best
and mature physical theories correspond to existing structures in the world
(as opposed to the Platonist \emph{ante rem}-picture of structures).
In the same sense should the quantum bundle structure be understood
as a physical \emph{in re}-structure of the world.

The most prominent example of \emph{in re}-structures that has been invoked
by structural realists so far is symmetry group structure
(cf. \npcite{kantorovich2003}, \npcite{lyre2004}).
Brian Roberts (\ynpcite{roberts2011}, p.~50) has aptly called
this position ``Group Structural Realism (GSR),'' and he defines it in the following way:
\emph{``The existing entities described by quantum theory are organized into a hierarchy,
in which a particular symmetry group occupies the top, most fundamental position.''}
Roberts adds that ``this statement of GSR should be taken as
a minimal assumption of the view'' allowing us ``to leave the exact nature
of a group structure's `existence' to the individual metaphysician.''
In the same spirit, we may broaden the scope of GSR by considering more
than just groups as physical structures.
Indeed, fundamental physics tells us that other structures are manifested as well.
Among the candidate structures that modern physics invites us to consider are,
most prominently, the external spacetime symmetry structure,
the internal gauge groups structure, and the quantum structure.

Structural realism is at the same time a moderate version of realism.
A full-blown realism about the quantum bundle would be a bundle space substantivalism.
But as in the debate about space-time ontology, where relationalists (anti-realists)
and substantivalists (realists) about spacetime stand opposed,
structural realism offers a third, intermediate route.
This fits nicely to the observation about Thomas precession in the foregoing section,
where it turned out that this effect is covered as long as one takes the Lorentzian structure
of spacetime into account -- which means, ontologically speaking,
that one should consider it a real trace of nature.
Minkowski spacetime is real insofar as it consists of a structure that constrains
the geodesic behavior of particles as well as the direction of gyroscopes.
Spacetime structuralism, unlike full-blown substantivalism, only takes this much
spacetime structure as real. Along the same lines, it is only the \emph{structure} of the quantum bundle
that I advocate to consider as real. As its most crucial structural feature
this structure entails the possibility of holonomies, which is just what the Berry phase is.

To conclude: Berry's quantum phase is neither brought about by the existence of forces
nor by the geometrical or topological properties of spacetime, but rather by
the geometrical structure of a U(1) fiber bundle over the projective Hilbert space.
And for precisely the same reasons why (structural) realists are well-advised
to consider the external space-time structure, the internal gauge group structure,
and the permutation symmetry structure of quantum theory as existing,
they are also advised to consider the quantum bundle structure as real.


\end{document}